\documentclass[lettersize,journal]{IEEEtran}
\usepackage{amsmath,amsfonts}
\usepackage{multirow} 
\usepackage{algorithmic}
\usepackage{array}
\usepackage[caption=false,font=normalsize,labelfont=sf,textfont=sf]{subfig}
\usepackage{textcomp}
\usepackage{stfloats}
\usepackage{url}
\usepackage{verbatim}
\usepackage{graphicx}
\usepackage{tabularx}
\usepackage{booktabs}
\def\BibTeX{{\rm B\kern-.05em{\sc i\kern-.025em b}\kern-.08em
    T\kern-.1667em\lower.7ex\hbox{E}\kern-.125emX}}
\usepackage{balance}
\begin{document}
\title{Robust Multi-Modal Speech In-Painting: A Sequence-to-Sequence Approach} 

\author{Mahsa Kadkhodaei Elyaderani$^{1}$, and Shahram Shirani$^{2}$%
\thanks{$^{1}$Mahsa Kadkhodaei Elyaderani is with the Department of Computational Science and Engineering, McMaster University, Hamilton, Canada (e-mail: kadkhodm@mcmaster.ca).}%
\thanks{$^{2}$Shahram Shirani is with the Department of Electrical and Computer Engineering, McMaster University, Hamilton, Canada (e-mail: shirani@mcmaster.ca).}}



\maketitle

\begin{abstract}
The process of reconstructing missing parts of speech audio from context is called speech in-painting. Human perception of speech is inherently multi-modal, involving both audio and visual (AV) cues. In this paper, we introduce and study a sequence-to-sequence (seq2seq) speech in-painting model that incorporates AV features. Our approach extends AV speech in-painting techniques to scenarios where both audio and visual data may be jointly corrupted. To achieve this, we employ a multi-modal training paradigm that boosts the robustness of our model across various conditions involving acoustic and visual distortions. This makes our distortion-aware model a plausible solution for real-world challenging environments.
We compare our method with existing transformer-based and recurrent neural network-based models, which attempt to reconstruct missing speech gaps ranging from a few milliseconds to over a second. Our experimental results demonstrate that our novel seq2seq architecture outperforms the state-of-the-art transformer solution by $38.8\%$ in terms of enhancing speech quality and $7.14\%$ in terms of improving speech intelligibility. We exploit a multi-task learning framework that simultaneously performs lip-reading (transcribing video components to text) while reconstructing missing parts of the associated speech.
\end{abstract}


\section{Introduction}

\IEEEPARstart{S}{peech} in-painting (SI) is restoring corrupted segments of speech signals caused by transmission issues, noisy environments, or damaged physical media. The purpose of SI models is to improve two perceptual aspects of distorted speech signals: quality and intelligibility. To ensure quality, the restored speech should sound natural, preserving the original speakers' prosody. For intelligibility, the content of the repaired speech should align coherently with intact segments and remain meaningful within the spoken language. An ideal SI method generalizes to unseen speakers and content variations without requiring additional information.

In audio in-painting, different durations of missing gaps are explored: short, medium, and long. Short gaps, lasting below $50$ $ms$, are commonly studied in many traditional audio in-painting approaches \cite{janssen, etter, adler}, often resulting from clicking corruptions in recording devices. Medium gaps, typically spanning above $50$ $ms$ and below $200$ $ms$, are usually caused by packet losses in transmissions. Methods like \cite{self, graph, contextenc} focus on recovering medium-length gaps. Long gaps (exceeding $200$ $ms$) and extra-long gaps (exceeding $1$ $s$) arise from severe connection issues, damaged media, or environmental noise during recording. However, only a few papers address the challenge of restoring long gaps, as demonstrated by \cite{Morrone, zhou}.

The earliest methods proposed for audio in-painting employ audio interpolation or extrapolation terms \cite{janssen}, \cite{etter}, \cite{wolfe}, \cite{oudre}, \cite{kauppinen}, \cite{esquef0}, and \cite{esquef1}. The auto-regressive (AR) model proposed in \cite{janssen} is one of the first methods in the area. Authors in \cite{janssen} apply an adaptive algorithm to estimate AR coefficients by minimizing the sum of squares of residual errors between corrupted and clean audio signal samples. But, the method in \cite{janssen} works with missing parts in audio that are short enough in length to be assumed stationary. As the stationary assumption of missing gaps limits AR-based models, the proposed method in \cite{etter} breaks down the AR parameters to the left and right sides of missing audio gaps. Hence, filled audio samples are the weighted left and right interpolations. Similarly, the work in \cite{kauppinen} utilizes forward and backward extrapolation for reconstructing missing audio segments with infinite impulse response (IIR) filter coefficients.

Introducing the term ``audio in-painting" for the first time, the method of \cite{adler} uses sparse representation modelling to approximate each audio frame in the time domain as a linear combination of the columns of a discrete cosine and Gabor dictionary. Another sparsity-based approach for audio in-painting is proposed in \cite{mokry}, employing the Alternating Direction Method of Multipliers (ADMM) \cite{ghadimi} for optimization. The notion of internal redundancies in audio pieces inspires the idea of filling missing parts with the most similar segments from a dataset. In \cite{self}, the proposed method uses self-similarity to in-paint audio gaps with the best match from available user records. Similarly, the sparse similarity graph introduced in \cite{graph} follows the same principle by concealing musical corruptions with the strongest connection in the defined graph. 

While methods such as those outlined in \cite{janssen, etter, adler, self, graph} focus on reconstructing short missing gaps of up to $50$ milliseconds, neural network-based approaches offer restoring longer missing audio segments. For instance, the encoder-decoder architecture method introduced in \cite{contextenc} employs convolutional and de-convolutional layers to reconstruct gap lengths of $64$ $ms$, utilizing reliable audio segments surrounding the missing part. Additionally, \cite{Kagler} proposes a U-net with a VGG-like deep feature extractor loss function called SpeechVGG, obtained by pre-training the renowned VGG model on classifying thousands of the most frequently spoken words in their training dataset. The work of \cite{Chang} casts the audio in-painting task into an image in-painting task via interpreting spectrograms as two-dimensional images. \cite{lin} proposes a complex U-net, incorporating convolutional layers with complex multiplication and addition operations to reconstruct spectrograms without discarding phase information. Similarly, \cite{Yu} and \cite{Chang} introduce convolutional models with either gated (for audio waveforms in-painting) or dilated/strided convolutions (for spectrograms in-painting) instead of standard convolutions. The method in \cite{masking} applies partial convolutions to ensure that the convolution of masked spectrograms depends only on uncorrupted pixels.  \cite{hasan} presents an integrated model of convolutional and recurrent networks with attention mechanisms to capture spectral and temporal features of audio spectrograms.\cite{khan} employs Cartesian Genetic Programming (CGP) to evolve parameters in an audio in-painting neural network. Additionally, \cite{cheddad} embeds Halftone-based Compression and Reconstruction (HCR) information into the Least-Significant-Bits (LSB) as side information, aiding the reconstruction of lost signals using Long Short-Term Memory Networks (LSTMs) and Random Forest (RF) models.

The perception of speech inherently involves both auditory and visual cues. AV perception is especially crucial for individuals with hearing impairments \cite{impaired}, vocal cord lesions, or in environments where acoustic signals are distorted or unreliable \cite{seeing}. A recent survey on visual speech analysis \cite{sheng} accentuates Visual Speech Recognition (VSR) or lip reading as integral components in speech enhancement and speech synthesis methods. Furthermore, studies \cite{hao, zhou2018visual, chen} have demonstrated that visual and auditory modalities are transformable into one another. Consequently, a branch of audio in-painting methods \cite{montesinos, zhou, Morrone, hou} employs visual features, such as hand movements, lip motions, and facial expressions, to recover audio distortions. To provide complementary information about audio signals, audio and visual modalities can be fused at various stages \cite{multimodalcombine}.

The proposed method of \cite{Morrone} incorporates facial landmark features from videos into the in-painting of speech spectrograms for gaps ranging from $100$ $ms$ to $1600$ $ms$ \cite{lstm}. The authors also demonstrate the utility of facial landmarks for speech enhancement in another study \cite{morroneface}. In the recent work, \cite{montesinos} leverages the audio-visual HuBERT network (AV-HuBERT) \cite{shi} as a pre-trained video encoder. Then, they temporally fuse visual encodings and acoustic features in an introduced transformer model. This transformer model \cite{montesinos} incorporates modality encoding alongside positional encodings. In a survey about multimodal learning (MML) \cite{mmlt}, various cross-modal interactions such as early summation, early concatenation, hierarchical attention, and cross-attention are identified as key practices for transformer-based multimodal modelling. Besides the HuBERT \cite{hubert} model, SpeechT5 \cite{speecht5}, Whisper \cite{whisper}, and Wav2vec 2.0 \cite{wav2vec} are transformer-based models for speech representation, applied across tasks including speech recognition, synthesis, and enhancement. Similar to \cite{montesinos}, \cite{borsos} exploits a transformer model named Perceiver IO \cite{jaegle} to synthesize missing speech content guided by text. Additionally, the text-informed model proposed in \cite{prablanc} converts text into speech and fills missing speech segments using synthesized audio from text.

Generative Neural Networks (GAN) \cite{goodfellow} have emerged as powerful models for audio synthesis, capable of generating high-fidelity raw audio samples \cite{mehri, oord, shen}. Among GAN models used for audio in-painting, \cite{gacela, wgan, zhou} have shown great success. In \cite{gacela}, authors introduce a conditional GAN that utilizes contextual information across multiple discriminators with different receptive fields to discern real spectrograms from fake ones. The Wasserstein GAN model presented in \cite{wgan} aims to restore missing audio from adjacent intact regions by minimizing the Wasserstein distance between ground truth and generated data distributions. The multi-modal GAN architecture outlined in \cite{zhou} provides a joint AV representation for audio in-painting. This architecture employs a convolutional encoder-decoder model to reconstruct missing audio disruptions of up to $800$ $ms$. Additionally, \cite{zhou} leverages the WaveNet generative model \cite{wavenet} to decode spectrogram outputs into high-quality audio waveforms. However, it's worth noting that the training process of GAN-based models for audio in-painting is computationally expensive.

While many audio in-painting methods \cite{janssen, etter, adler, self, Kagler, Chang, contextenc} focus on music or environmental sound datasets, our paper introduces a novel method for speech in-painting. Our approach comprises an encoder and a decoder. The encoder takes visual features, while the decoder utilizes fused audio-visual features to in-paint Mel-spectrogram representations within a seq2seq model \cite{seq2seq}. The contributions of our methods are three-fold:

\begin{itemize}
 \item Firstly, our seq2seq model is the first audio-visual speech in-painter that attempts to reconstruct missing speech segments when both audio and video modalities are corrupted. As far as our knowledge extends, no prior model has tackled this challenge. We achieve this by using a multi-modal data augmentation training paradigm.
 
\item Secondly, our training paradigm enables us to incorporate speech enhancement into speech in-painting. Experiments show that our proposed approach enhances noise robustness without requiring a noise-suppressing module. 

\item Finally, using a novel seq2seq architecture and a multi-task learning framework, we significantly enhance the state-of-the-art model for speech in-painting \cite{montesinos} over the Grid Corpus. Our approach improves speech quality by $38.8\%$ and speech intelligibility by $7.14\%$. Notably, we achieve these results via less than $10\%$ of the number of the trainable parameters compared to the transformer-based model \cite{montesinos}.
\end{itemize}

The remainder of this paper is organized as follows. Section \ref{sec:method} introduces the proposed multi-modal seq2seq model, loss functions, and a brief explanation of our pre- and post-processing steps, which will be expanded in Section \ref{sec:exp} with details on the implementations. Section \ref{sec:exp} defines all the competing models and the evaluation metrics used in this paper. Our experimental results in speech in-painting, speech enhancement, and sentence predictions under various visual and acoustic conditions are presented in Section \ref{sec:res}. Section \ref{sec:res} shows how our proposed model outperforms the state-of-the-art speech in-painting transformer. Lastly, Section \ref{sec:con} concludes the paper and sketches our future directions.

\section{Our Proposed Model}
\label{sec:method}
\begin{figure}[t]
\centering
\begin{picture}(350,245)
\put(0,0){\includegraphics[width=1\columnwidth]{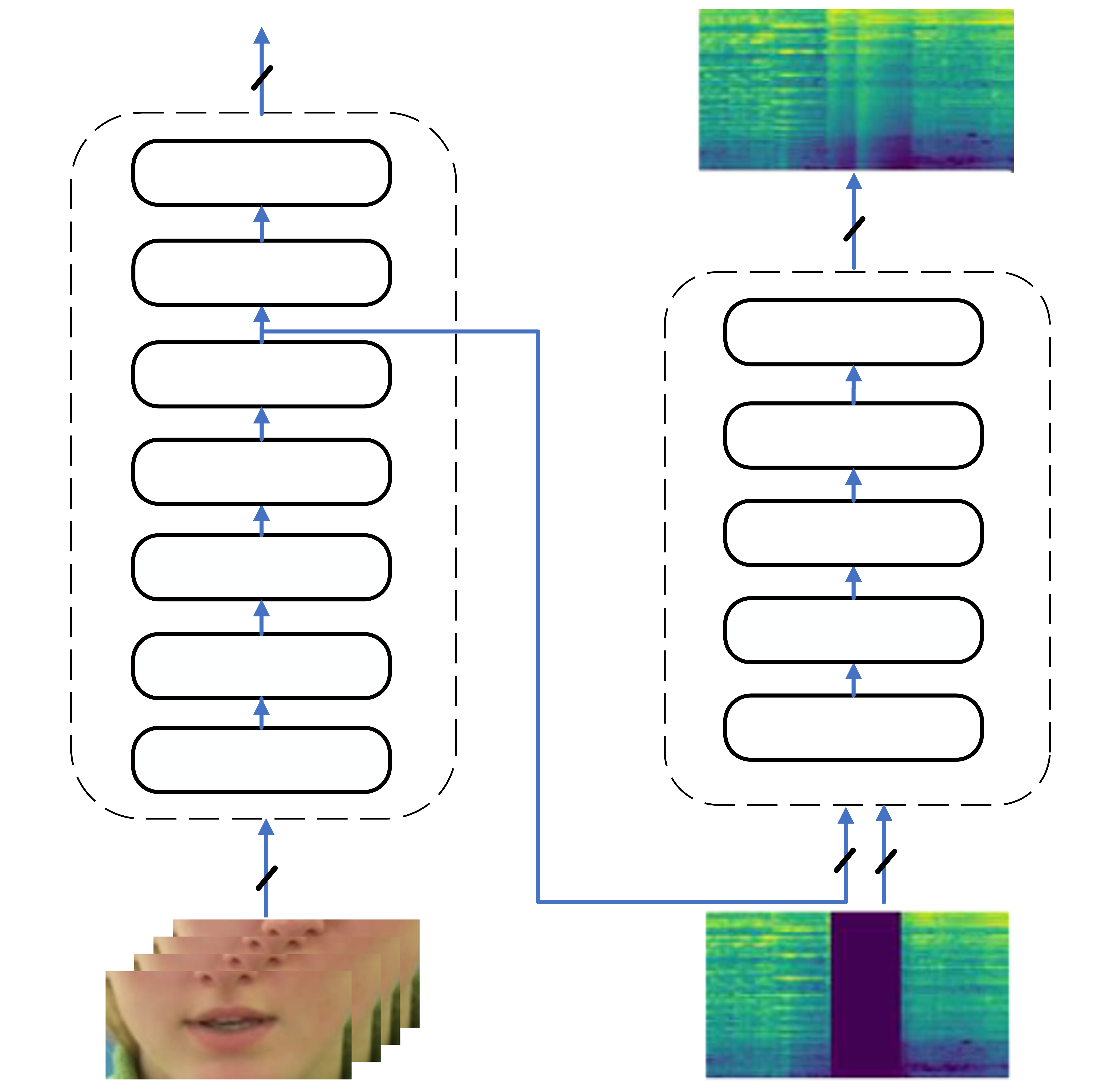}}

\put(62,227){$[p_{1},\cdots, p_{T^\prime}]$}
\put(63,48){$[v_1,\cdots, v_{T^\prime}]$}
\put(195,193){$[y_1,\cdots, y_T]$}
\put(202,49){$[a_1,\cdots, a_T]$}
\put(136,49){$[c_1,\cdots, c_{T^\prime}]$}

\put(10,242){``Place red at G9 now."}
\put(41,204){Softmax}
\put(51,181){FC}
\put(40,158){BLSTM}
\put(40,136){BLSTM}
\put(40,115){Conv3D}
\put(40, 92){Conv3D}
\put(40,72){Conv3D}
\put(5,70){\rotatebox{90}{\textbf{Encoder}}}

\put(187,168){FC}
\put(175,144){BLSTM}
\put(175, 123){BLSTM}
\put(175,101){BLSTM}
\put(176,79){Concat}
\put(240,70){\rotatebox{90}{\textbf{Decoder}}}
\end{picture}
 \caption{An illustration of the seq2seq model for speech in-painting proposed in this paper. The encoder or the lip-reader (on the left) takes cropped video frames and outputs corresponding transcriptions. The decoder or the speech in-painter (on the right) uses both spectral and visual features from the encoder to restore distorted spectrograms.}
  \label{fig:diag}
\end{figure}
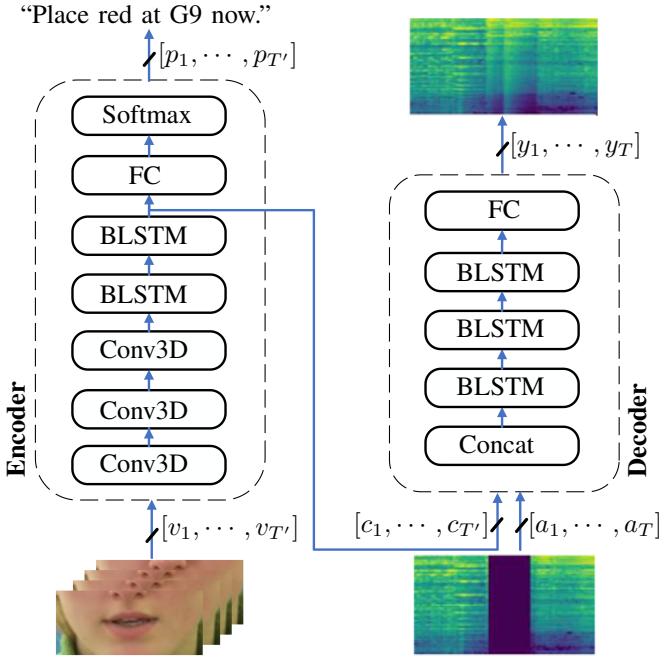

In this section, we introduce our seq2seq model for speech in-painting with AV data. The proposed model consists of an encoder and a decoder for visual and auditory modalities, respectively. Considering the sequential nature of speech, we design an architecture based on Recurrent Neural Networks (RNN), specifically Bidirectional LSTM (BLSTM) networks, to accomplish this task end-to-end. Figure \ref{fig:diag} illustrates the overview of our proposed method, which will be explained in detail throughout the rest of this section.

\subsection{Pre-Processing Stage}

Given $[ x_1, x_2, \cdots, x_T]$ as the clean Mel-spectrogram representation of a speech signal, where $x_t$ for $1\le t \le T$ represents the frequency vector corresponding to time $t$, the corrupted Mel-spectrogram at each time step $a_t$ is obtained by element-wise multiplication of $x_t$ with the binary mask $m_t \in { 0, 1 }$, denoted as $a_t = m_t \cdot x_t$.

Similarly, video frames undergo face detection to detect frontal faces. Then, a landmark feature extractor \cite{dlib} is utilized to draw facial landmark points from the detected faces. These facial landmark points are used to crop the video frames, retaining only the mouth regions. The resulting cropped video frames are inputs to the encoder, denoted as $[v_1, v_2, \cdots, v_{T^\prime}]$. 

\subsection{Encoder}
The left box of Figure~\ref{fig:diag} is the encoder of our model, which is similar to the LipNet model \cite{lipnet}. It leverages spatiotemporal convolutions (STCNN), BLSTMs, and the connectionist temporal classification (CTC) loss to translate a variable-length sequence of video frames into text. The architecture of the encoder contains three STCNNs, with dropout and max-pooling layers in between. Then, the extracted features are fed into two BLSTM layers. Subsequently, a non-linear transformation, followed by a softmax activation function is applied. 
We use the formulation:
\begin{align}
f_{enc}(V; \theta_{enc}) & = \mathbb{P} (p_{1}, p_{2}, \cdots, p_{T^\prime} |v_{1}, v_{2}, \cdots , v_{T^\prime}), 
\end{align}
 where $\theta_{enc}$ represents the encoder's trainable parameters, $V=[v_1, v_2, \cdots, v_{T^\prime}]$ denotes the sequence of video frames, and $[p_1, p_2, \cdots, p_{T^\prime}]$ signifies the sequence of character probabilities.
 
The motivation for training the encoder as a lip reader is the correlation between lip movements and spoken characters. We hypothesize that the features extracted from the top BLSTM layer of the encoder offer valuable cues for the decoder to fill in the missing segments of the audio signal.
 
\subsection{Decoder}
In our pipeline, the right box illustrated in Figure~\ref{fig:diag} represents the decoder, tasked with restoring the missing segments of the Mel-spectrograms. The decoder not only takes the Mel-spectrogram features of the distorted input audio, denoted by $A = [a_1, a_2, \cdots, a_T]$, but also incorporates the outputs of the last BLSTM layer of the encoder, represented by $[c_1, c_2, \cdots, c_{T^\prime}]$. These features are temporally concatenated and then fed into the decoder to reconstruct the corrupted parts of the Mel-spectrograms. The decoder architecture consists of three BLSTM layers and a fully connected (FC) layer to compute the following expression:
\begin{align}
	\label{eq:lstm2}
      f_{dec}(A, C; \theta_{dec}) = \mathbb{P} (y_{1}, y_{2}, \cdots, y_{T} | a_{1}, a_{2}, 	\cdots, a_{T}, \nonumber \\
      c_{1}, c_{2}, \cdots, c_{T^{\prime}}).   
 \end{align}
 
Here, $\theta_{\text{dec}}$ represents the trainable parameters of the decoder, $[y_1, y_2, \cdots, y_T]$ denotes the decoder's output, and $A = [a_{1}, a_{2}, \cdots, a_{T}]$ and $C = [c_{1}, c_{2}, \cdots, c_{T^{\prime}}]$ signify the input Mel-spectrogram and the output of the encoder's top BLSTM layer, respectively. Essentially, the decoder estimates the conditional probability of the output given the Mel-spectrogram input and the encoder's contextual features.

The mean squared error (MSE) loss function minimizes the difference between the predicted Mel-spectrograms, i.e. $[y_1, y_2, \cdots, y_T]$, and the ground truth Mel-spectrograms.

\subsection{Post-Processing Stage}
Following the model prediction, the reconstructed Mel-spectrogram is completed differently depending on whether it's an "informed" or "uninformed" speech in-painting method. In the case of informed speech in-painting, the locations of the corrupted parts of Mel-spectrograms are known. Hence, frequency vectors of a Mel-spectrogram are reconstructed as $$o_t = m_t \cdot x_t + (1 - m_t) \cdot y_t,$$ where $[y_1, y_2, \cdots, y_T]$ represents the outputs of the decoder, $[m_1, m_2, \cdots, m_T]$ indicates the mask information (i.e., corrupted locations in Mel-spectrograms), and $\cdot$ denotes element-wise multiplication between sequences.

However, in the uninformed speech in-painting, the reconstruction simplifies to:
$$o_t = y_t, \hspace{0.2 in} 1\le t \le T.$$

 Then, we convert the output Mel-spectrogram, $o_t$, into a linear spectrogram. Subsequently, the Griffin-Lim algorithm \cite{griffin} iteratively restores the phase of spectrograms and applies an inverse STFT to attain audio signals.
 
\subsection{Loss Function}
\label{loss}
The proposed model, depicted in Figure~\ref{fig:diag}, has a hybrid loss function comprising the CTC and the MSE terms for the encoder and decoder, respectively. The CTC loss function encourages the model to produce sequences that maximize the probability of aligning with the target sequence, while it penalizes misalignments between the predicted and target sequences. This is mathematically expressed as:
\begin{align}
    \mathcal{L}_\text{\textit{enc}} = \arg \min_{\theta_{enc}} - \Pi_{t=1}^{T^\prime} 	\mathbb{P}(z_{t} | p_{t}; \theta_{enc}),
\end{align}
where $z_{t}$ and $p_{t}$ represent the target and encoder's output characters at time step $t$, respectively.
Also, the MSE is defined as the average squared difference between the target and decoder's output spectrograms as in Equation~\ref{eq:mse}.
\begin{align}
\label{eq:mse}
    \mathcal{L}_\text{\textit{dec}} = \frac{1}{T} \sum_{t=1}^{T}(x_t - y_t)^2.
\end{align}

Here, $X=[ x_1, x_2, \cdots, x_T]$ and $Y=[y_{1}, y_{2}, \cdots, y_{T}]$ are the target and decoder's output spectrograms, respectively. 

The hybrid loss is a weighted sum of the two aforementioned losses similar to Equation~\ref{eq:loss}. 
\begin{equation}
\label{eq:loss}
\mathcal{L}_\text{\textit{total}}= \mathcal{L}_\text{\textit{dec}} + \lambda \cdot \mathcal{L}_\text{\textit{enc}} 
\end{equation}
In this expression, $\lambda$ is a trade-off parameter. The trade-off parameter balances each of our defined loss functions for two separate tasks.

\section{Experiments}
\label{sec:exp}

\subsection{Data Preparation}
\label{sec:data_prepare}

We perform our experiments over the Grid Corpus \cite{grid}, which comprises $34$ speakers, $18$ males and $16$ females. The speakers utter $1000$ sentences, each lasting $3$ seconds. Besides audio signals and video frames, transcripts aligned with the spoken sentences are available. The sentences in this corpus have a predefined structure consisting of a command, colour, preposition, letter, digit, and adverb. The command words are selected from the set \{bin, lay, place, set\}, while colours are chosen from \{blue, green, red, white\}. The prepositions are drawn from \{at, by, in, with\}, and the adverbs from \{again, now, please, soon\}. The letters and digits range from \{A, $\cdots$, Z\} (excluding W) and \{zero, $\cdots$, nine\}, respectively.

\subsubsection{Dataset}

In our study, the train and test split of the Grid Corpus is based on speakers. Specifically, our training set comprises $29$ speakers, totaling $25k$ samples, denoted as $s1-20$, $s22-29$, and $s31$. Meanwhile, the testing sets consist of $4$ speakers. Speaker $s32-34$, $s30$ for the unseen test set ($3k$ samples), and $s26-27$, $s29$, and $s31$ for the seen test set ($2k$ samples). Samples from speakers $s26-27$, $s29$, and $s31$ are evenly distributed between the train and seen test sets to create an overlapped test set. The speaker $s33$ is excluded from all the sets because of incomplete data. 

\subsubsection{Audio data}

We convert audio speech signals into Mel-spectrogram representations. Initially, audio signals are down-sampled from $25$ $kHz$ to $8$ $kHz$ to make the large audio dimensions more manageable, and a pre-emphasis filter is applied to preserve high frequencies of audio signals. Subsequently, the Short-Time Fourier Transform (STFT) is performed on each audio signal, keeping only the magnitude of the STFT. We employ the STFT using windows of $320$ sample points (equivalent to $40$ $ms$) and hop lengths of $160$ points (equivalent to $20$ $ms$). During the STFT computation, the length of the windowed signals, after zero-padding, becomes $510$ sample points. Then, Mel filter banks are applied, followed by log transformation. Mel-scaled spectrograms contain $64$ frequency bins and $149$ temporal units. Finally, Mel-spectrograms are normalized to ensure that inputs fall within the range of $0$ to $1$. To introduce missing audio segments in the Mel-spectrograms, we randomly draw durations from a normal distribution with a mean of $900$ $ms$ and a standard deviation of $300$ $ms$. These durations are uniformly split into $1$ to $8$ gaps, each with a minimum length of $36$ $ms$.

\subsubsection{video data}

The videos have the original sampling rates of $25$ $fps$. For video preprocessing, we use the DLib face detector \cite{dlib} to capture frontal faces in video frames. Then, we use the iBug face shape predictor \cite{ibug} to extract $68$ facial landmarks from the detected faces and crop the mouth regions based on the extracted facial landmarks.

\subsubsection{Data Augmentation}
\label{sec:data_aug}

Among effective techniques for improving model generalizability, data augmentation stands out by enriching the diversity of the training dataset. We augment our AV dataset using affine transformations to generate new examples while preserving the ground truths. We incorporate various types of noise, including white noise and environmental sounds like airplane landings, heavy rain, dog barking, and car noises, into the Mel-spectrograms. This mitigates the limitations of models trained solely on clean recordings. Additionally, we apply masks to video frames to induce synchronized audio and video distortions. These video masks are equal to those used for Mel-spectrograms but are downsampled to match the video frame rate.

\subsection{Competing Methods}
\label{subsec:comp}
Here is the list of models we compare in this paper:

\subsubsection{Audio Speech In-painting \normalfont(\textbf{A-SI})}
This is the audio-only baseline model that uses masked spectrograms to reconstruct clean spectrograms. Its architecture resembles the decoder part of our model in Figure~\ref{fig:diag}, comprising three BLSTM layers and one FC layer. 

\subsubsection{AV Speech In-painting \normalfont(\textbf{AV-SI})}
This model, from \cite{Morrone}, represents the RNN-based state-of-the-art. It comprises only three BLSTM layers and one FC layer, lacking an encoder-decoder architecture. We replicate their method by temporally concatenating synchronized motion vectors of video frames and Mel-spectrograms. However, our implementation differs in that we extract motion vectors from mouth regions of consecutive frames, whereas \cite{Morrone} utilizes landmark points from entire faces. Additionally, we utilize MSE loss, contrasting with \cite{Morrone}'s use of $L_1$ loss.

\subsubsection{AV Multi-Task Speech In-painting \normalfont(\textbf{AV-MTL-SI})}
This is another model from \cite{Morrone}, with a similar architecture to \textbf{AV-SI}. While in-painting speech, this model predicts spoken sentences using the CTC loss function discussed in Section \ref{loss}.

\subsubsection{AV Transformer \normalfont(\textbf{AV-T-SI})}
This model is the transformer-based state-of-the-art from \cite{montesinos}. The visual features are initially extracted from intact video frames using the AV-HuBERT \cite{avhubert}. Then, Mel-spectrograms and visual features are concatenated by a positional encoding, and a modality encoding to be fed into the proposed speech in-painting transformer.

\subsubsection{AV Seq2Seq \normalfont(\textbf{AV-S2S})}
Similar to Figure \ref{fig:diag}, this model has an encoder-decoder architecture \cite{me}. It takes motion vectors between facial landmarks of consecutive frames as encoder inputs. Then, the encoder top LSTM layer outputs are concatenated with audio spectrograms for the decoder inputs.

\subsubsection{AV Multi-Task Seq2Seq \normalfont(\textbf{AV-MTL-S2S})}
This is another solution in \cite{me}, sharing similarities with \textbf{AV-S2S} in architecture and inputs. However, its loss function is designed to in-paint Mel-spectrograms and predict the spoken sentences, as illustrated in Equation \ref{eq:loss}.

\subsubsection{AV Multi-Task Convolutional Seq2Seq \normalfont(\textbf{AV-MTL-CS2S})}
This is the proposed model shown in Figure~\ref{fig:diag}. As explained in Section~\ref{sec:method}, it takes cropped video frames as the encoder's inputs. The decoder's inputs consist of concatenated audio Mel-spectrograms and the top BLSTM layer output from the encoder. The encoder is inspired by the lip-reading model proposed in \cite{lipnet}. Moving forward from multi-modal speech in-painters that rely on visual feature extraction methods, we suggest a progression towards an end-to-end architecture for AV speech in-painting.

We develop an augmented training paradigm by introducing audio and visual distortion to the training set, as mentioned in \ref{sec:data_prepare}. The models trained with the augmented train set have names starting with \textbf{AUG-}, e.g. \textbf{AUG-AV-S2S}.

\subsection{Evaluation Metrics}
To evaluate our models' performance, we use several metrics comparing predictions with ground truths in both frequency (Mel-spectrograms) and time domains (waveforms). The Perceptual Evaluation of Speech Quality (PESQ) \cite{pesq} and Short Term Objective Intelligibility (STOI) \cite{stoi} measure audio quality and intelligibility by comparing reference and reconstructed audio signals, respectively. The Peak Signal-to-Noise Ratio (PSNR) and Mean Squared Error (MSE) assess the quality and fidelity of reconstructed Mel-spectrograms. The PESQ, STOI, and PSNR evaluate the entire audio signal or Mel-spectrogram representation, while MSE focuses solely on errors in the missing segments.
Character Error Rate (CER) and Word Error Rate (WER) evaluate the generated text in multi-task models. CER measures the minimum single-character edits needed to match predicted sequences to original sentences using the Levenshtein distance \cite{edit}. WER uses the same approach for words.

\subsection{Implementation Setup}
We use Adam optimization \cite{adam} with a learning rate of $0.0001$ for \textbf{AV-MTL-CS2S} and \textbf{AUG-AV-MTL-CS2S}, and $0.001$ for other models. Default hyperparameters are employed for momentum coefficients ($0.9$ and $0.999$) and the numerical stability parameter ($\epsilon = 10^{-8}$). Network parameters are initialized using He initialization \cite{he}, except for BLSTM layers, which are initialized orthogonally \cite{chung}. ReLU activation is applied to all dense and convolutional layers, while BLSTM layers utilize the $tanh$ function. Also, the top FC layer of the encoder uses the Softmax function. Table~\ref{tab:enc-fltr} outlines the model hyperparameters, optimized through random search. Each BLSTM layer has a latent dimensionality of $256$. The dimension order of convolutional layers is $temporal \times height \times width \times channels$. The Repeat layer upsamples the output of the top BLSTM layer in the encoder, enabling temporal concatenation with Mel-spectrograms in the next layer.

\begin{table}[b]
   \caption{Model architecture hyperparameters.}
     \label{tab:enc-fltr}
  \centering
   \begin{tabular}{c c c}
    \hline
    \textbf{Layers} & \textbf{Size/Stride}& 
    \textbf{Input Size}\\
    \hline

    STCNN & $3\times5\times5 / 1, 2, 2$ & $75 \times 50 \times 100 \times 3$\\
    Pool & $1\times2\times2 / 1, 2, 2 $ & $75 \times 25 \times 50 \times 128$\\
    STCNN & $3\times5\times5 / 1, 2, 2 $ & $75 \times 12 \times 25 \times 128$\\
    Pool & $1\times2\times2 / 1, 2, 2 $ & $75 \times 12 \times 25 \times 256$\\
    STCNN & $3\times3\times3 / 1, 2, 2 $ & $75 \times 6 \times 12 \times 75$\\
    Pool & $1\times2\times2 / 1, 2, 2 $ & $75 \times 6 \times 12 \times 75$\\
    BLSTM & $256$ & $75 \times (3 \times 6 \times 75)$\\
    BLSTM & $256$ & $75 \times 512$\\
    Dense & $256$ & $75 \times 512$\\
    Softmax & $41$ & $75 \times 256$\\
     
     \hline
     \hline
     
    Repeat & - & $75 \times 512 $\\
    Concatenate & - & $(149 \times 512) + (149 \times 64) $\\
    BLSTM & $256$ & $149 \times (512 + 64)$\\
    BLSTM & $256$ & $149 \times 512$\\
    BLSTM & $256$ & $149 \times 512$\\
    Dense & $64$ & $149 \times 512$\\
    \hline
  \end{tabular}
\end{table}
In multi-task learning models, we leverage the loss function from Equation \ref{eq:loss} with a trade-off parameter $\lambda$ set to $0.001$. For other models, we use MSE. The tuned learning rate decays after five consecutive epochs without improvement for all models except \textbf{AUG-AV-MTL-CS2S}, where it decays after one epoch without improvement. The models' training stops if there are no significant improvements for four consecutive epochs for \textbf{AUG-AV-MTL-CS2S}, or twenty epochs for the others. Mini-batch sizes are $2$ for \textbf{AV-MTL-CS2S} and \textbf{AUG-AV-MTL-CS2S}, and $32$ for the rest.
The total number of trainable parameters in \textbf{AV-MTL-CS2S} is $12M$. In comparison, the models proposed in \cite{montesinos, me, Morrone} have complexities of $300M$, $9M$, and $4M$ parameters, respectively. For all experiments, we report the average of specified metrics over all the test set samples.

All models are deployed on the TensorFlow Keras $2.11.0$ backend \cite{keras}. We exploit the Librosa Python package for processing speech signals \cite{librosa}. The implementations run on the NVIDIA GeForce RTX 4090 GPU with 24 GB RAM and the 13\normalfont$^{th}$-generation Intel Core i9 CPU.

\section{Results and Discussions}
\label{sec:res}
Tables \ref{tab:informed} and \ref{tab:uninformed} compare the performance of our \textbf{AV-MTL-CS2S} model against competing methods using the dataset without distortions described in Section \ref{sec:data_prepare}. Additionally, Table \ref{tab:aug} presents the performance of all models on the augmented dataset, which includes both auditory and visual distortions. For further insight, Tables \ref{tab:align} and \ref{tab:align-aug} show the text prediction errors of the multi-task learning models with and without distortions, respectively. The following subsections will provide detailed discussions of conducted experiments.

\subsection{Quantitative Evaluation}

\begin{table*}[tbp]
    \caption{A comparison of the informed speech in-painting methods studied in this paper in terms of STOI, PESQ, PSNR, and MSE. Upward arrows indicate higher values are better, while lower values are better for downward arrows.}
    \label{tab:informed}
  \centering
  \resizebox{.99\textwidth}{!}{
   \begin{tabular}{c c c c c c c c c}
   \hline
    \multirow{2}{*}{\textbf{Methods}} & 
    \multicolumn{4}{c}{\textbf{Unseen Speakers}} & 
    \multicolumn{4}{c}{\textbf{Seen Speakers}} \\
    \cmidrule(lr){2-5} \cmidrule(lr){6-9}
     & 
    \textbf{PESQ}$^{\uparrow}$ & 
    \textbf{STOI}$^{\uparrow}$ & 
     \textbf{PSNR}$^{\uparrow}$& 
     \textbf{MSE}$^{\downarrow}$&
     \textbf{PESQ}$^{\uparrow}$ & 
     \textbf{STOI}$^{\uparrow}$ & 
     \textbf{PSNR}$^{\uparrow}$& 
     \textbf{MSE}$^{\downarrow}$\\
     \hline
     Input                       & 1.59 & 0.63 & 13.97 & 0.17 & 1.56 & 0.63 & 13.81 & 0.18~~~\\
     A-SI             & 2.23 & 0.78 & 25.52 & 0.01 & 2.25 & 0.78 & 25.64 & 0.01~~~\\
     AV-SI \cite{Morrone}        & 2.30 & 0.80 & 25.85 & 0.01 & 2.41 & 0.82 & 26.37 & 0.01~~~\\
     AV-MTL-SI \cite{Morrone}        & 2.04 & 0.76 & 25.06 & 0.02 & 2.08 & 0.76 & 25.08 & 0.01~~~\\
     AV-T-SI \cite{montesinos} & 2.21 & 0.84 & - & - & - & - & - & -~~~\\
     AV-S2S \cite{me} & 2.32  & 0.81 & 25.83 & 0.01 & 2.48 & 0.83 & 26.58 & 0.01~~~\\
     AV-MTL-S2S \cite{me} & 2.33 & 0.80 & 25.84 & 0.01 & 2.38 & 0.81 & 26.23 & 0.01~~~\\
     AV-MTL-CS2S & \textbf{3.07} & \textbf{0.90} & \textbf{32.96} & \textbf{0.004} & \textbf{3.25} & \textbf{0.90} & \textbf{34.01} & \textbf{0.003}~~~\\
     \hline
  \end{tabular}}
\end{table*}

\begin{table*}[tbp]
    \caption{A comparison of the uninformed speech in-painting methods studied in this paper in terms of STOI, PESQ, PSNR, and MSE. Upward arrows indicate higher values are better, while lower values are better for downward arrows.}
    \label{tab:uninformed}
  \centering
  \resizebox{.99\textwidth}{!}{
   \begin{tabular}{c c c c c c c c c }
   \hline
    \multirow{2}{*}{\textbf{Methods}} & 
    \multicolumn{4}{c}{\textbf{Unseen Speakers}} & 
    \multicolumn{4}{c}{\textbf{Seen Speakers}} \\
    \cmidrule(lr){2-5} \cmidrule(lr){6-9}
    \textbf{} & 
    \textbf{PESQ}$^{\uparrow}$ & 
    \textbf{STOI}$^{\uparrow}$ & 
     \textbf{PSNR}$^{\uparrow}$& 
     \textbf{MSE}$^{\downarrow}$&
     \textbf{PESQ}$^{\uparrow}$ & 
     \textbf{STOI}$^{\uparrow}$ & 
     \textbf{PSNR}$^{\uparrow}$& 
     \textbf{MSE}$^{\downarrow}$\\
     \hline
     Input                       & 1.59 & 0.63 & 13.97 & 0.17 & 1.56 & 0.63 & 13.81 & 0.18~~~\\
     A-SI             & 2.12  & 0.76 & 24.85 & 0.02 & 2.14 & 0.76 & 25.06 & 0.01~~~\\
     AV-SI \cite{Morrone}        & 2.22  & 0.79 & 25.48 & 0.01 & 2.34 & 0.80 & 25.98 & 0.01~~~\\
     AV-MTL-SI \cite{Morrone}        & 1.36  & 0.64 & 22.49 & 0.02 & 1.53 & 0.65 & 22.58 & 0.01~~~\\
     AV-S2S \cite{me} & 2.27  &  0.80 & 25.69 & 0.01 & 2.44 & 0.82 & 26.47 & 0.01~~~\\
     AV-MTL-S2S \cite{me} & 2.17 & 0.78 & 25.13 & 0.01 & 2.26 & 0.78 & 25.49 & 0.01~~~\\
     AV-MTL-CS2S & \textbf{2.27} & \textbf{0.80} & \textbf{25.94} & \textbf{0.004} & \textbf{2.48} & \textbf{0.82} & \textbf{27.19} & \textbf{0.003}~~~\\
     \hline
  \end{tabular}}
\end{table*}

The first rows of Tables~\ref{tab:informed} and \ref{tab:uninformed} compare unprocessed masked Mel-spectrogram inputs with ground truths. The \textbf{A-SI} model notably enhances unprocessed inputs. Introducing landmark visual feature motions in \textbf{AV-SI} and \textbf{AV-S2S} outperforms \textbf{A-SI}, with our seq2seq model displaying superior performance in both informed and uninformed cases. This highlights the potential of seq2seq architectures in speech-processing applications. For simulated corruptions within the range of $300$ $ms$ to $1500$ $ms$, the \textbf{AV-MTL-S2S} model surpasses \textbf{AV-MTL-SI} across all evaluation metrics. Additionally, in Table~\ref{tab:informed}, the \textbf{AV-MTL-S2S} model yields comparable results to \textbf{AV-S2S} for unseen speakers, demonstrating the \textbf{AV-MTL-S2S} model's capabilities in both speech in-painting and text prediction. Yet, \textbf{AV-S2S} shows higher quality and intelligibility than \textbf{AV-MTL-S2S} when in-painting missing gaps for seen speakers in both informed and uninformed cases, as well as for unseen speakers in the uninformed case.

Our proposed \textbf{AV-MTL-CS2S} model exceeds all models in the informed case by a significant margin. It achieves outstanding results compared to the \textbf{AV-T-SI} model with ten times fewer trainable parameters. Evaluating the MSE of all models reveals that \textbf{AV-MTL-CS2S} excels in re-generating missing parts of speech spectrograms. The \textbf{AV-MTL-S2S} achieves the second lowest MSE across all experiments, indicating the effectiveness of multi-task learning in minimizing the MSE of missing speech signals. However, the \textbf{AV-MTL-CS2S} model has comparable results to \textbf{AV-S2S} for unseen speakers in the uninformed case. We hypothesize that incorporating a mask prediction mechanism similar to \cite{masking} can potentially elevate the performance of \textbf{AV-MTL-CS2S} and \textbf{AV-MTL-S2S}, as demonstrated by test results for the informed case.

The \textbf{AV-MTL-SI} model exhibits the poorest performance among all models, suggesting that multi-task learning is less suitable for this architecture. Overall, all metrics are higher for seen speakers compared to unseen ones.

\subsubsection{Effect of gap sizes}
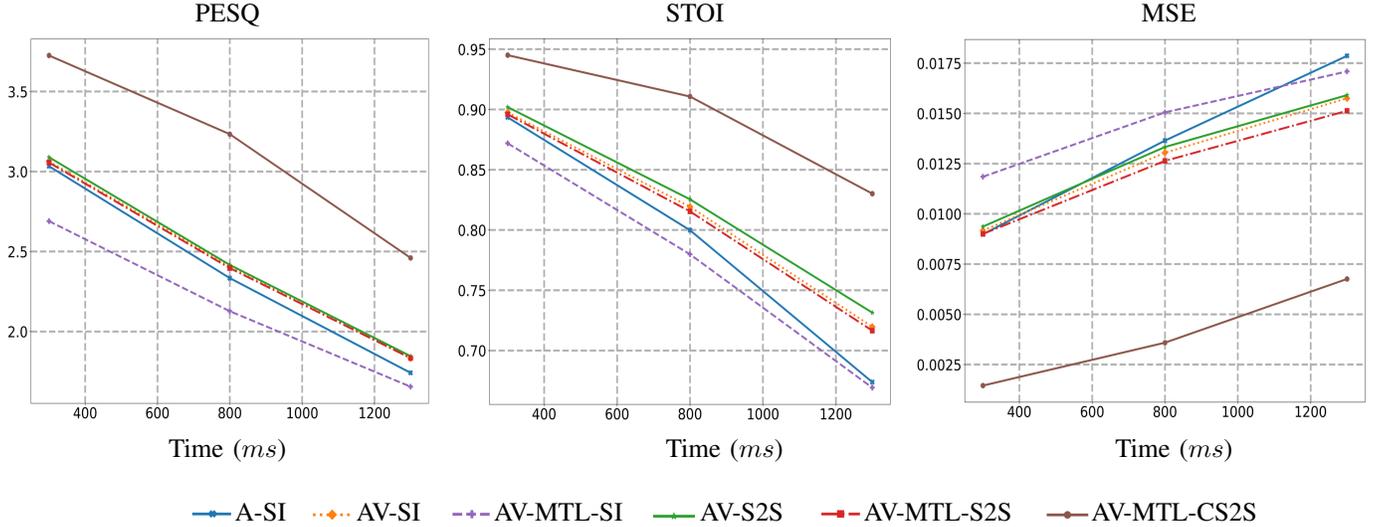
\begin{figure*}[tbp]
\centering
\caption{Effects of gap sizes on the reconstructed audio quality and intelligibility plus spectrograms quality in the informed case for the unseen speakers' test set. The horizontal axis of all diagrams shows the duration of gaps in milliseconds, and the vertical axis is the models' performances in terms of each evaluation metric.}
\label{fig: info-gaps}
\begin{picture}(520,210)
\put(0,0){\includegraphics[width=1\textwidth]{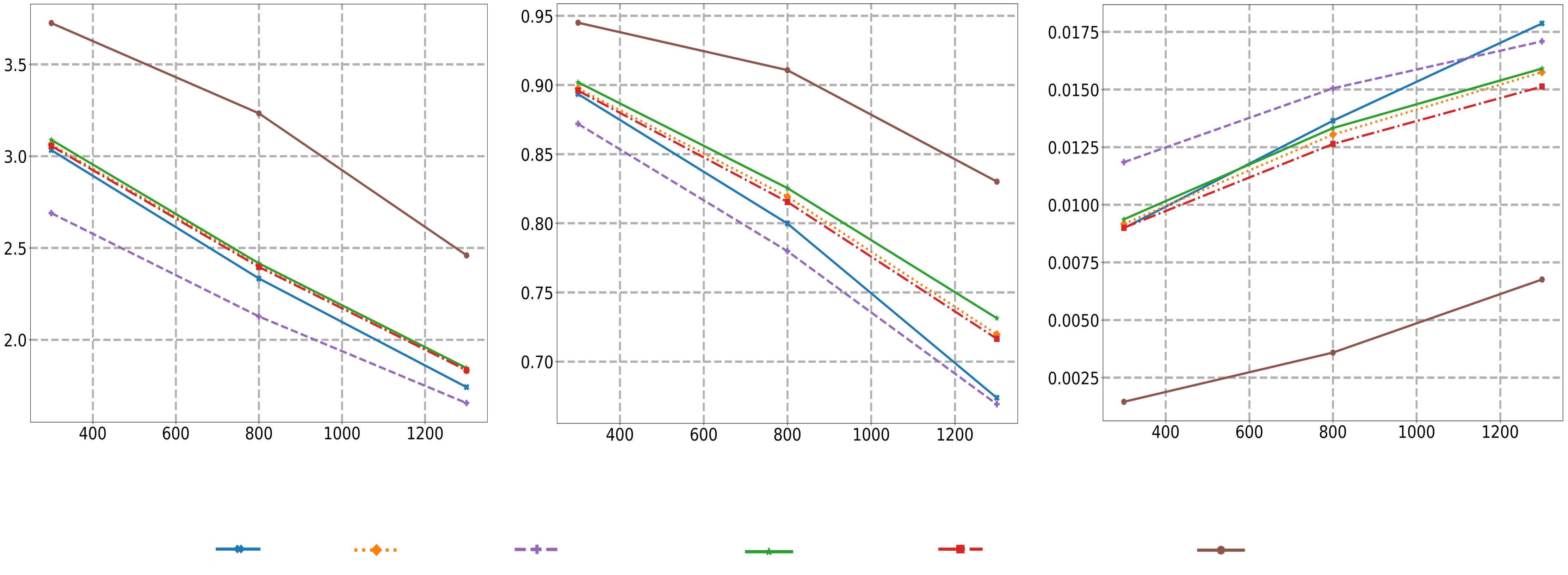}}

\put(72,195){PESQ}
\put(62, 30){Time ($ms$)}

\put(250,195){STOI}
\put(250, 30){Time ($ms$)}

\put(430, 195){MSE}
\put(420, 30){Time ($ms$)}

\put(87, 6){A-SI}
\put(133, 6){AV-SI}
\put(185, 6){AV-MTL-SI}
\put(263, 6){AV-S2S}
\put(325, 6){AV-MTL-S2S}
\put(411, 6){AV-MTL-CS2S}

\end{picture}
\end{figure*}

In Tables~\ref{tab:informed} and \ref{tab:uninformed}, we compare all models against missing gaps that are randomly drawn from a normal distribution with a mean of $900$ $ms$ and a standard deviation of $300$ $ms$. Figure~\ref{fig: info-gaps} illustrates the performance of models across three gap ranges—short, long, and extra-long—that are derived from normal distributions with means of $300$ $ms$, $800$ $ms$, and $1300$ $ms$, respectively, and a standard deviation of $300$ $ms$. The outputs of all models are evaluated on the unseen speakers' test set and reconstructed in the informed case. The horizontal axis represents gap durations in milliseconds, while the vertical axis denotes model performance across evaluation metrics.

 In Figure~\ref{fig: info-gaps}, \textbf{AV-S2S}, \textbf{AV-SI}, and \textbf{AV-MTL-S2S} are very similar in audio quality as shown in PESQ. The \textbf{AV-MTL-CS2S} maintains the best performance within all gap ranges and evaluating metrics. The significant difference in terms of the MSE makes us believe that multi-task learning is effective in minimizing the error of reconstructing gaps. This can be observed in the MSE of the \textbf{AV-MTL-S2S}, being the lowest after the \textbf{AV-MTL-CS2S}.
 Among all competing methods, \textbf{AV-MTL-SI} has the lowest performance, except for very long gaps, longer than 1 second, in which the \textbf{AV-MTL-SI} beats the \textbf{A-SI} model in the MSE metric. 

\subsubsection{Audio-Visual Distortions}

\begin{table*}[tbp]
    \caption{A comparison of the speech in-painting models on the contaminated unseen speakers' test set in the uninformed case. Before adding acoustic noises, inputs of the unseen test set had $PESQ=1.59$, $STOI=0.63$, $PSNR=13.97$, and $MSE=0.17$. Upward arrows indicate higher values are better, while lower values are better for downward arrows.}
    \label{tab:aug}
  \centering
  \resizebox{.99\textwidth}{!}{
   \begin{tabular}{c c c c c c c c c}
    \hline
    \multirow{2}{*}{\textbf{Methods}} & 
    \multicolumn{4}{c}{\textbf{Acoustic Distortions}} & 
    \multicolumn{4}{c}{\textbf{Visual Distortions}} \\
    \cmidrule(lr){2-5} \cmidrule(lr){6-9}
    \textbf{} & 
    \textbf{PESQ}$^{\uparrow}$ & 
    \textbf{STOI}$^{\uparrow}$ & 
     \textbf{PSNR}$^{\uparrow}$& 
     \textbf{MSE}$^{\downarrow}$&
     \textbf{PESQ}$^{\uparrow}$ & 
     \textbf{STOI}$^{\uparrow}$ & 
     \textbf{PSNR}$^{\uparrow}$& 
     \textbf{MSE}$^{\downarrow}$\\
     \hline
     Noisy Input                       & 1.16 & 0.45 & 10.33 & 0.17 & 1.59 & 0.63 & 13.97 & 0.17~~~\\
     A-SI          & 1.37 & 0.57 & 17.21 & 0.03 & - & - & - & - ~~~\\
     AV-SI \cite{Morrone} & 1.41 & 0.60 & 17.67 & 0.03 & 2.06 & 0.73 & 22.85 & 0.03~~~\\
     AV-MTL-SI \cite{Morrone}   & 1.27  & 0.55 & 18.06 & 0.02 &1.38 & 0.61 & 21.48 & 0.02~~~\\
     AV-S2S \cite{me} & 1.34 & 0.58 & 17.22 & 0.03 & 2.02 & 0.72 & 22.42 & 0.04~~~\\
     AV-MTL-S2S \cite{me} & 1.35 & 0.57 & 17.29 & 0.03 & 1.98 & 0.72 & 23.34 & 0.02~~~\\
     AV-MTL-CS2S & 1.54 & 0.58 & 15.67 & 0.09 & 1.93 & 0.70 & 22.37 & 0.01~~~\\
     
     \hline
     
     AUG-A-SI          & 1.35  & 0.56 & 17.17 & 0.04 & - & - & - & -~~~\\
     AUG-AV-SI \cite{Morrone} & 1.39  & 0.58  & 18.29 & 0.03 & 2.15 & 0.77  & 24.46 &0.02~~~\\
     AUG-AV-MTL-SI \cite{Morrone}   & 1.28 & 0.55 & 18.07  & 0.02 & 1.38 & 0.61& 21.48 & 0.02~~~\\
     AUG-AV-S2S \cite{me} & 1.34 & 0.59 & 17.35 & 0.03 & 2.13 & 0.77 & 23.97 & 0.02~~~\\
     AUG-AV-MTL-S2S \cite{me} & 1.42 & 0.60 & 13.37 & 0.02 & 2.04 & 0.74 & 23.51 & 0.02~~~\\
     AUG-AV-MTL-CS2S & \textbf{2.13} & \textbf{0.78} & \textbf{25.52} & \textbf{0.004} & \textbf{2.20} & \textbf{0.77} & \textbf{25.57} & \textbf{0.005}~~~\\
     \hline
  \end{tabular}}
\end{table*}
To evaluate the models' performance in imperfect environments, we subject them to additive white noise, environmental sounds, and missing visual features. We chose these two types of acoustic noises because environmental sounds, such as airplane or rain sounds, have underlying dependencies, while white noise is uncorrelated in time. Additive white noise follows a normal distribution with a mean of zero and a standard deviation of one, and it is added to Mel-spectrograms with a magnitude of $0.1$. Additionally, environmental sound has a magnitude of $0.05$.

Table~\ref{tab:aug} presents our models' speech in-painting performance on the unseen test set contaminated with AV distortions. All Mel-spectrograms in Table~\ref{tab:aug} are reconstructed in the uninformed case. Models trained with the augmented train set are denoted by names beginning with \textbf{AUG}. In our analysis, we evaluate each model's capacity to suppress acoustic noises while in-painting missing speech segments. The introduced gaps in Table~\ref{tab:aug} align with those in Tables~\ref{tab:informed} and \ref{tab:uninformed}, with a mean of $900$ $ms$ and a standard deviation of $300$ $ms$. The first row in Table~\ref{tab:aug} reflects the quality and intelligibility of unprocessed test sets after the injection of each distortion.

Among the models, the \textbf{AUG-AV-MTL-CS2S} model outperforms the others in acoustic distortions, followed by \textbf{AV-MTL-CS2S}. Significantly, \textbf{AV-MTL-CS2S} exhibits even better results than \textbf{AUG-AV-S2S} and \textbf{AUG-AV-MTL-S2S}, which were trained with the augmented dataset. However, all other models, including \textbf{AV-MTL-S2S}, \textbf{AV-S2S}, \textbf{AV-MTL-SI}, and \textbf{AV-SI}, yield results similar to the audio-only model. This suggests that our proposed augmented AV model with multi-task learning excels in real acoustic environments.

According to Table~\ref{tab:aug}, visual distortions have a lesser impact on all audio metrics than acoustic distortions. Nevertheless, \textbf{AUG-AV-MTL-CS2S} demonstrates the best performance in reconstructing high-quality audio signals, followed by \textbf{AUG-AV-S2S}. Notably, among the non-augmented models, those without multi-task learning, i.e. \textbf{AV-SI} and \textbf{AV-S2S}, achieve the best results. This can be attributed to the additional effort required during training for text predictions in the presence of unreliable visual and acoustic information, potentially leading to lower speech in-painting quality and intelligibility for these models. In summary, Table~\ref{tab:aug} highlights that our \textbf{AUG-AV-MTL-CS2S} model demonstrates greater robustness than \textbf{AV-SI} and \textbf{AV-MTL-SI} for the speech in-painting task.

\subsubsection{Alignment}
Table~\ref{tab:align} provides a detailed comparison of sentence predictions, in terms of CER and WER for the \textbf{AV-MTL-CS2S}, \textbf{AV-MTL-S2S}, and \textbf{AV-MTL-SI} models. We improve our previous work \textbf{AV-MTL-S2S} \cite{me} in text prediction which can show the robustness of our extracted visual features against landmark features. The \textbf{AV-MTL-CS2S} model shows similar errors to the \textbf{AV-MTL-SI} model with unseen speakers but shows fewer errors with seen speakers. Based on some manual inspection that we conducted, it seems that the differences between CER and WER in our model are primarily due to misspelled word predictions such as “son”, “thre”, and “gren” instead of the correct words “soon”, “three”, and “green”. As a result, we incorporate a corrective enhancement (CE) for spell-checking from the Grid Corpus dictionary. The revised results, which include spell correction, are indicated by the suffix \textbf{+CE} added to the model names. Notably, the \textbf{AV-MTL-CS2S+CE} model depicts significant improvements for unseen speakers in terms of WER compared to the \textbf{AV-MTL-SI+CE} model.

\begin{table}[tbp]
  \caption{A comparison of the text prediction methods studied on the unseen speakers' test set in terms of WER and CER. Lower values are better for downward arrows.}
  \label{tab:align}
  \centering
   \begin{tabular}{c c c c c}
    \hline
    \multirow{2}{*}{\textbf{Methods}} &
    \multicolumn{2}{c}{\textbf{Unseen Speakers}}& 
    \multicolumn{2}{c}{\textbf{Seen Speakers}}\\
    \cmidrule(lr){2-3} \cmidrule(lr){4-5}
    \textbf{} & \textbf{CER}$^{\downarrow}$ & \textbf{WER}$^{\downarrow}$ &
                \textbf{CER}$^{\downarrow}$ & \textbf{WER}$^{\downarrow}$\\
    \hline
     AV-MTL-SI \cite{Morrone} & \textbf{10\%} & \textbf{26\%} & 8\% & 21\%~~~\\
     AV-MTL-S2S \cite{me} & 18\% & 44\% & 22\% & 48\%~~~\\
     AV-MTL-CS2S & 13\% & 30\% & \textbf{6\%} & \textbf{18\%}~~~\\
     \hline
     AV-MTL-SI + CE \cite{Morrone} & \textbf{9\%} & \textbf{18\%} & 7\% & 14\%~~~\\
     AV-MTL-S2S + CE \cite{me} & 20\% & 34\% & 24\% & 39\%~~~\\
     AV-MTL-CS2S + CE & 10\% & 20\% & \textbf{4\%} & \textbf{10\%}~~~\\
     \hline
  \end{tabular}
\end{table}

Table~\ref{tab:align-aug} investigates the robustness of multi-task learning models against acoustic and visual distortions in sentence predictions. We have also applied the corrective enhancement step to the results in Table~\ref{tab:align-aug}. Here, our \textbf{AUG-AV-MTL-CS2S} and \textbf{AUG-AV-MTL-CS2S+CE} models have matched error rates with the \textbf{AUG-AV-MTL-SI} and \textbf{AUG-AV-MTL-SI+CE} models for acoustic distortions. 

\begin{table}[tbp]
   \caption{A comparison of the text prediction methods studied on the unseen speakers' test set with acoustic and visual distortions in terms of WER and CER. Lower values are better for downward arrows.}
     \label{tab:align-aug}
  \centering
   \begin{tabularx}{0.5\textwidth}{c c c c c}
    \hline
    \multirow{2}{*}{\textbf{Methods}} &
    \multicolumn{2}{c}{\textbf{Acoustic Distortion}}& 
    \multicolumn{2}{c}{\textbf{Visual Distortion}}\\
    \cmidrule(lr){2-3} \cmidrule(lr){4-5}
    \textbf{} & \textbf{CER}$^{\downarrow}$ & \textbf{WER}$^{\downarrow}$ &
                \textbf{CER}$^{\downarrow}$ & \textbf{WER}$^{\downarrow}$\\
    \hline
     AUG-AV-MTL-SI \cite{Morrone} & \textbf{13\%} & \textbf{32\%} & \textbf{17\%} & \textbf{39\%}\\
     AUG-AV-MTL-S2S \cite{me} & 17\% & 40\% & 53\% & 87\%\\
     AUG-AV-MTL-CS2S & 14\% & 34\% & 21\% & 44\%\\
     \hline
     AUG-AV-MTL-SI + CE \cite{Morrone} & 13\% & 25\% & \textbf{16\%} & \textbf{30\%}\\
     AUG-AV-MTL-S2S + CE \cite{me} & 16\% & 30\% & 55\% & 74\%\\
     AUG-AV-MTL-CS2S + CE & \textbf{13\%} &  \textbf{24\%} & 21\% & 35\%\\
     \hline
  \end{tabularx}
\end{table}

Even though our \textbf{AV-MTL-CS2S} model is $50 \%$ better than \textbf{AV-MTL-SI} \cite{Morrone} in terms of speech in-painting metrics under all test conditions, it falls short of improving \textbf{AV-MTL-SI} \cite{Morrone} in terms of CER and WER metrics. We believe that we can bridge the gap in terms of CER and WER by further hyper-parameter tuning and slightly sacrificing speech in-painting improvements of \textbf{AV-MTL-CS2S}. Since the main focus of our work is on speech enhancement, conducting such an experiment is out of the scope of this paper. The WER and CER results are only reported for completeness.

\subsection{Qualitative Evaluation}

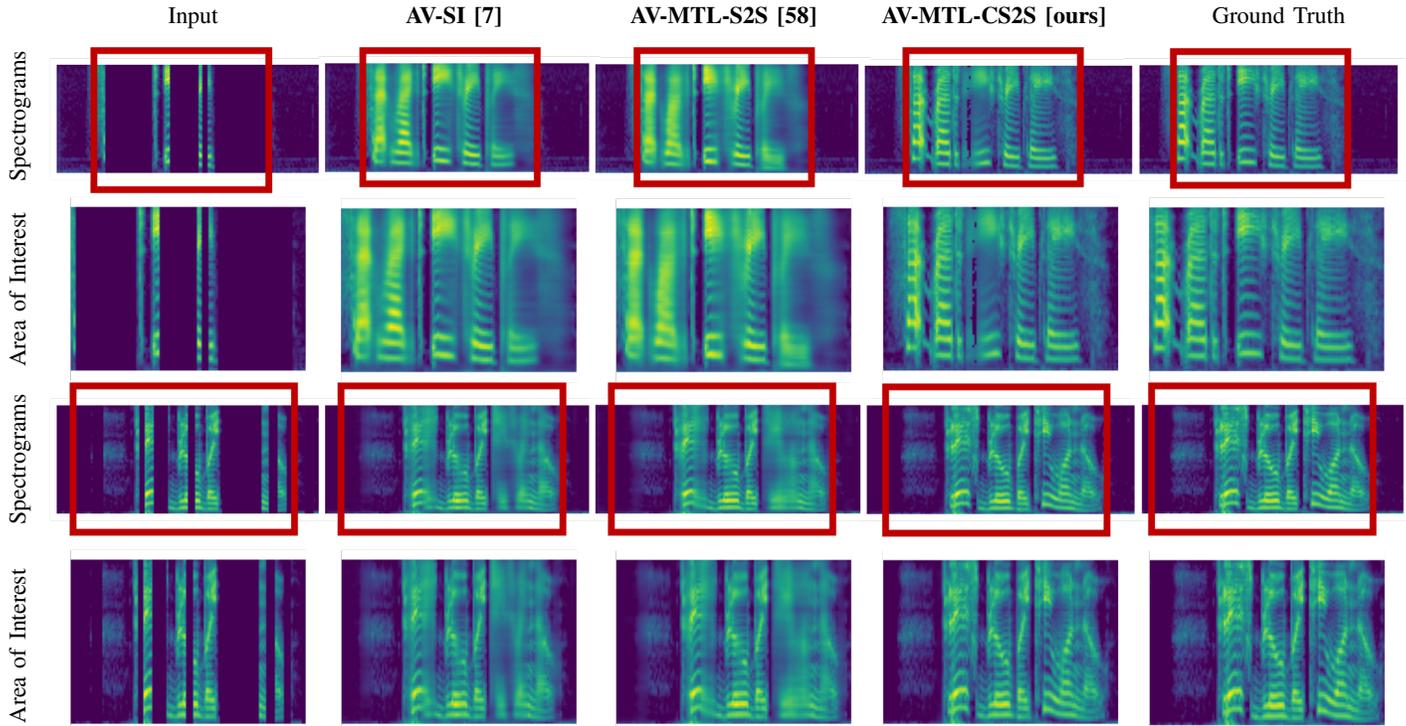
\begin{figure*}[tbp]
\centering
\begin{picture}(500,270)
\put(0,0){\includegraphics[width=1\textwidth]{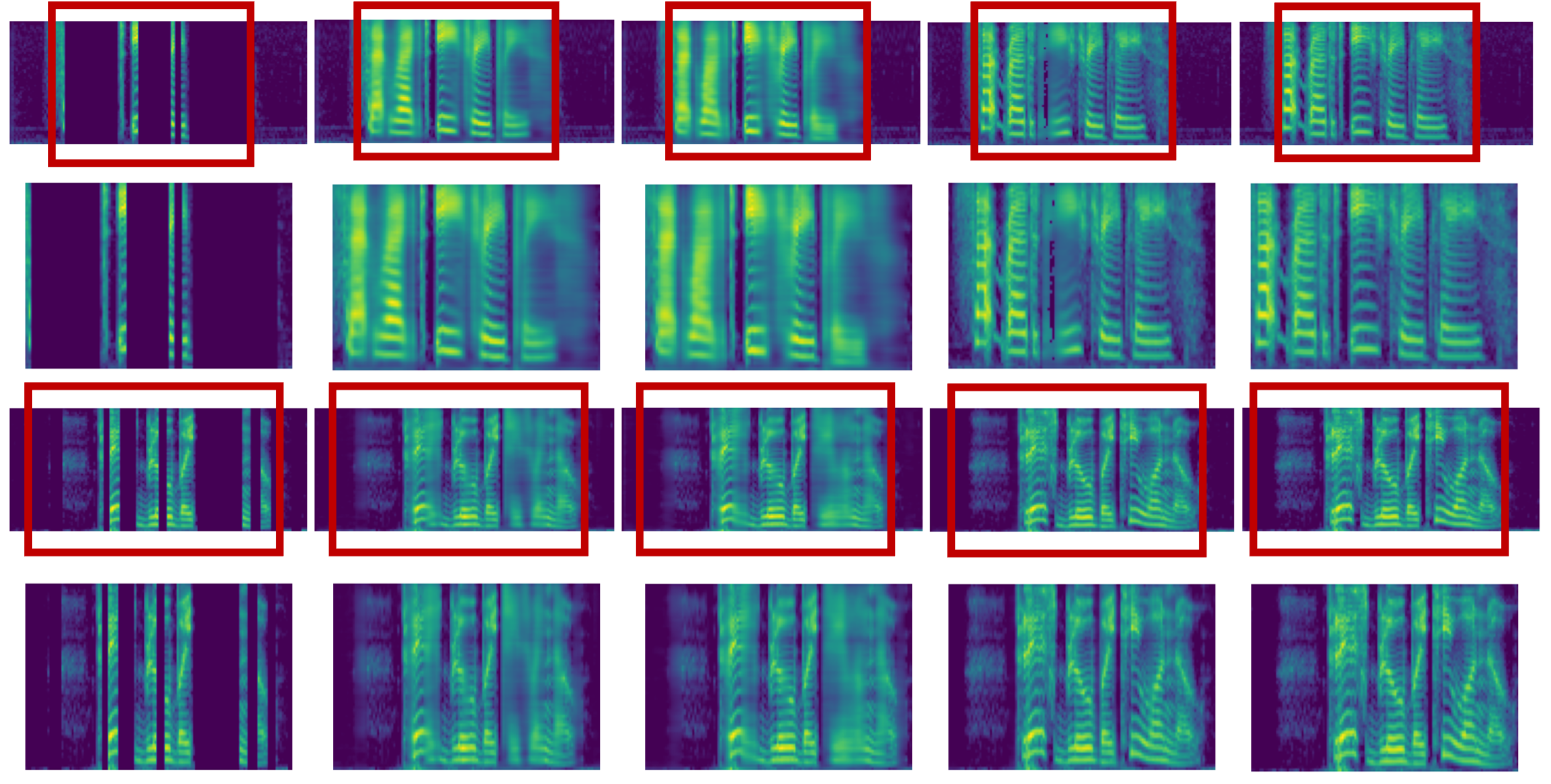}} 
\put(45,270){\small Input}
\put(135,270){\small \textbf{AV-SI \cite{Morrone}}}
\put(220,270){\small \textbf{AV-MTL-S2S \cite{me}}}
\put(315,270){\small \textbf{AV-MTL-CS2S [ours]}}
\put(440,270){\small Ground Truth}

\put(-15,210){\rotatebox{90}{\small Spectrograms}}
\put(-15,140){\rotatebox{90}{\small Area of Interest}}
\put(-15,80){\rotatebox{90}{\small Spectrograms}}
\put(-15,5){\rotatebox{90}{\small Area of Interest}}
\end{picture}
\caption{Qualitative results of in-painted Mel-spectrograms for \textbf{AV-SI \cite{Morrone}}, \textbf{AV-MTL-S2S \cite{me}}, and \textbf{AV-MTL-CS2S} models. The distorted areas are the areas of interest in red boxes and are zoomed in for better visualization. The first two rows exhibit the first example in the informed case, and the last two are the second example in the uninformed case.}
\label{fig:result}
\end{figure*}

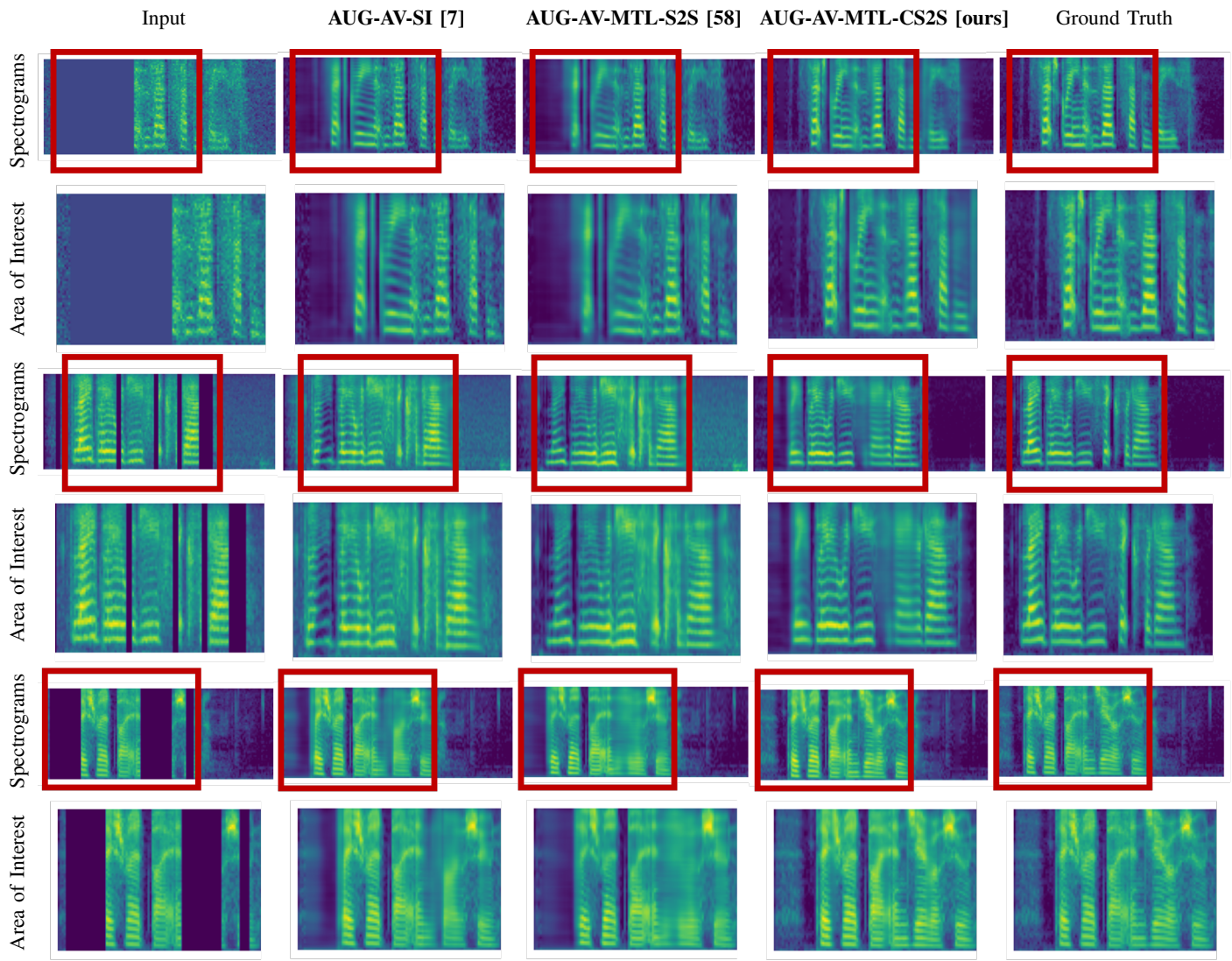
\begin{figure*}[tbp]
\centering
\begin{picture}(500,410)
\put(0,0){\includegraphics[width=1\textwidth]{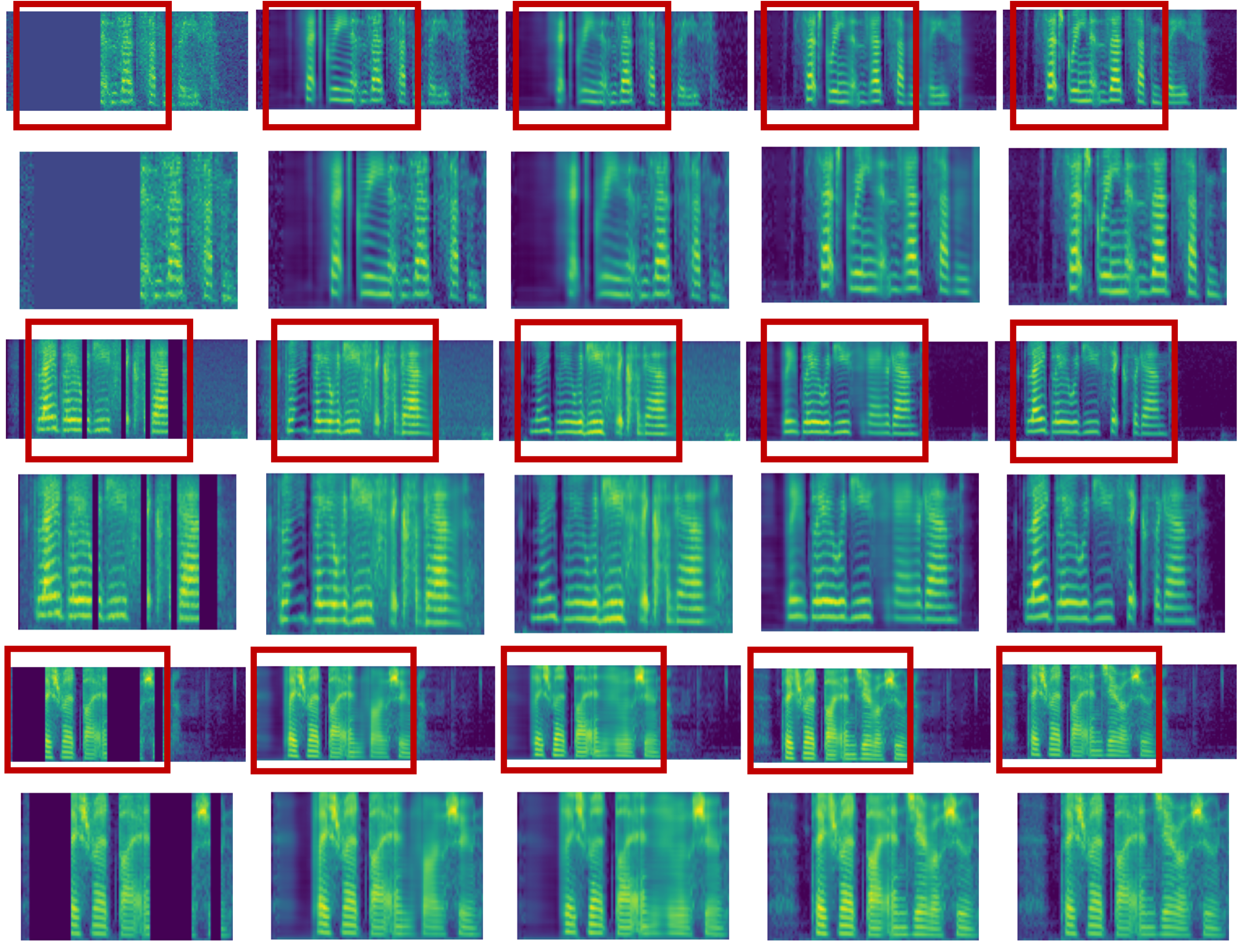}}

\put(45,405){\small Input}
\put(122,405){ \small \textbf{AUG-AV-SI \cite{Morrone}}}
\put(211,405){\small \textbf{AUG-AV-MTL-S2S \cite{me}}}
\put(312,405){\small \textbf{AUG-AV-MTL-CS2S [ours]}}
\put(440,405){\small Ground Truth}

\put(-12,343){\rotatebox{90}{\small Spectrograms}}
\put(-12,270){\rotatebox{90}{\small Area of Interest}}
\put(-12,210){\rotatebox{90}{\small Spectrograms}}
\put(-12,138){\rotatebox{90}{\small Area of Interest}}
\put(-12,75){\rotatebox{90}{\small Spectrograms}}
\put(-12,5){\rotatebox{90}{\small Area of Interest}}

\end{picture}
\caption{Qualitative results of in-painted Mel-spectrograms for \textbf{AUG-AV-SI \cite{Morrone}}, \textbf{AUG-AV-MTL-S2S \cite{me}}, and \textbf{AUG-AV-MTL-CS2S} models. The distorted areas are the areas of interest in red boxes and are zoomed in for better visualization. The first two rows exhibit the first example with additive white noise, the third and fourth rows are the second example with added environmental sounds, and the last two rows have synchronous masked AV data.}
\label{fig:result-noise}
\end{figure*}

Figures~\ref{fig:result} and \ref{fig:result-noise} show the reconstructed spectrograms from three models detailed in this study. The first and last columns present the distorted inputs and target Mel-spectrograms, respectively. The second, third, and fourth columns demonstrate the outcomes of employing \textbf{AV-SI} \cite{Morrone}, \textbf{AV-MTL-S2S} \cite{me}, and \textbf{AV-MTL-CS2S} on two distorted audio samples, respectively. Corrupted regions in the Mel-spectrograms are highlighted by red bounding boxes, with subsequent rows zooming into these areas. In Figure~\ref{fig:result}, the top row corresponds to the unseen test data, generated under the informed case, while the third-row data comes from the seen test set, reconstructed under the uninformed case. The resultant images reveal that the \textbf{AV-MTL-CS2S} model outperforms \textbf{AV-SI} \cite{Morrone} and \textbf{AV-MTL-S2S} \cite{me} in capturing textural complexities. Comparing \textbf{AV-SI} \cite{Morrone} with \textbf{AV-MTL-S2S} \cite{me}, they display similar visual characteristics, with some blurry reconstructions in missing audio regions.

Figure~\ref{fig:result-noise} illustrates the results of \textbf{AUG-AV-SI} \cite{Morrone}, \textbf{AUG-AV-MTL-S2S} \cite{me}, and \textbf{AUG-AV-MTL-CS2S} when dealing with corrupted Mel-spectrograms containing background noises, either additive white noise or environmental sounds, or missing visual information. In Figure~\ref{fig:result-noise}, the first row corresponds to inputs with additive white noise, the third row represents inputs with added environmental sounds, and the fifth row has synchronous masked audio and video segments. Notably, \textbf{AUG-AV-MTL-CS2S} shows higher visual quality in reconstructing and enhancing inputs simultaneously when compared with the ground truth. Both \textbf{AUG-AV-MTL-S2S} \cite{me} and \textbf{AUG-AV-SI} \cite{Morrone} have poor denoising performance which is visible in silent and speaking regions of generated Mel-spectrograms. However, \textbf{AUG-AV-MTL-CS2S} generates visually reasonable and continuous results, particularly with the additive white noise, indicating that environmental sounds and simultaneous modality corruptions pose greater challenges.

\section{Conclusion}
\label{sec:con}
In this paper, we introduced a robust seq2seq speech in-painting model. Leveraging a multimodal training paradigm with data augmentation, our proposed model reconstructs missing speech segments in scenarios where both audio and video modalities are corrupted. Additionally, our model suppresses acoustic noise while performing the speech in-painting task. We further enhanced our approach by incorporating multi-task learning to predict spoken sentences while generating spectral features. We evaluated our model on a speech dataset to reconstruct missing words or word pieces of durations ranging from short ($100$ $ms$) to extra long ($1600$ $ms$). Through several experiments, we demonstrated that incorporating visual features significantly enhanced the quality and intelligibility of reconstructed audio signals. Notably, we achieved state-of-the-art results in speech in-painting on the Grid Corpus with only 12 million training parameters.

In future studies, we will evaluate model performance by varying noise levels. In this paper, we kept the magnitude of acoustic noise constant during the experiments to compare all models under identical conditions. Additionally, we plan to work on multi-speaker datasets to evaluate whether multi-modal transformer-based models justify the cost of replacing RNN networks in multi-speaker settings.

\bibliographystyle{IEEEtran}
\bibliography{ref}

\end{document}